\documentclass{emulateapj}
\usepackage{apjfonts}

\begin{document}

\title{Testing Bimodal Planet Formation}

\author{Dale L. Fields and Andrew Gould}
\affil{Department of Astronomy, The Ohio State University,
140 W.\ 18th Ave., Columbus, OH 43210}
\email
{fields,gould@astronomy.ohio-state.edu}


\begin{abstract}
We suggest that the observed break in giant-planet frequency as a function of
host metallicity at $Z=0.02$ may be a
reflection of bimodal planet formation.  We search for signatures of
this bimodality in the distributions of the planet eccentricities,
periods, masses, and multiplicity.  However, the low-metallicity sample is at
present too small to test for any but the most severe differences
in these two putative populations.
\end{abstract}


\section{Test of Bimodality}

\begin{figure}[b]
\centerline{\epsfxsize=3.5truein\epsffile{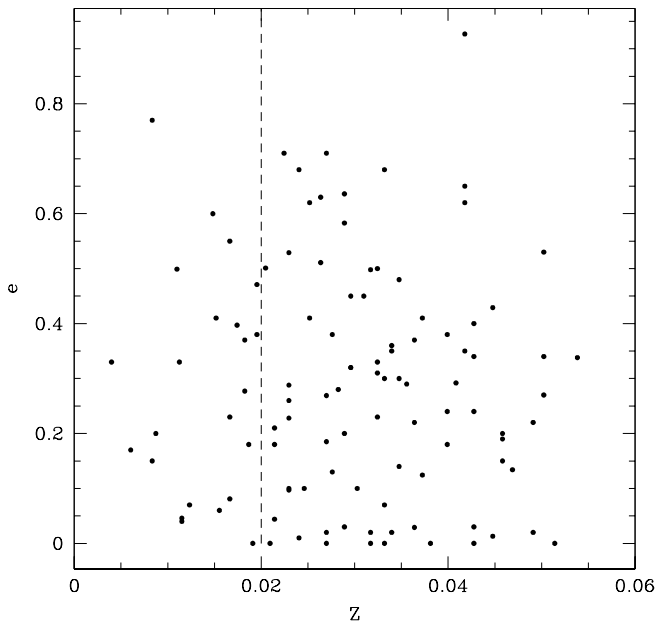}}
\caption{\label{fig:eccen}
Eccentricity $e$ vs.\ metallicity $Z$ for the 109 planets of the 98
stars with metallicities determined by \citet{santos} with the exception
of HD 47536b.  If the break in the frequency of planets at $Z=0.02$
reflects bimodal planet formation, one might expect this to be reflected
in a similar break in the eccentricity distribution at this boundary
({\it dashed line}).  No such break is visible, but the present
statistics would be sufficient to reveal only the most glaring
differences.  }\end{figure}

\citet{fischer} have demonstrated that the frequency of 
extra-solar giant planets is a strong function of metallicity $Z$.
\citet*{santos} have confirmed this result and have further 
shown that there is a sharp break in frequency at $Z=0.02$, which
can be represented algebraically by,
\begin{equation}
f = 0.025 + 16(Z-0.02)\Theta(Z-0.02),
\label{eqn:frequency}
\end{equation}
where $\Theta$ is a step function.  See their figure 7.  We suggest 
that the two terms of this equation correspond to two different modes of
giant-planet formation, the first being metallicity-independent and the
second being strongly dependent on metallicity.  For example, in its
simplest form, the disk-instability model of \citet{boss} would not be
expected to depend on metallicity.  By contrast, the more conventional
picture of gas accretion onto rock-ice cores might well be very
sensitive to metallicity.

\begin{figure}[b]
\centerline{\epsfxsize=3.5truein\epsffile{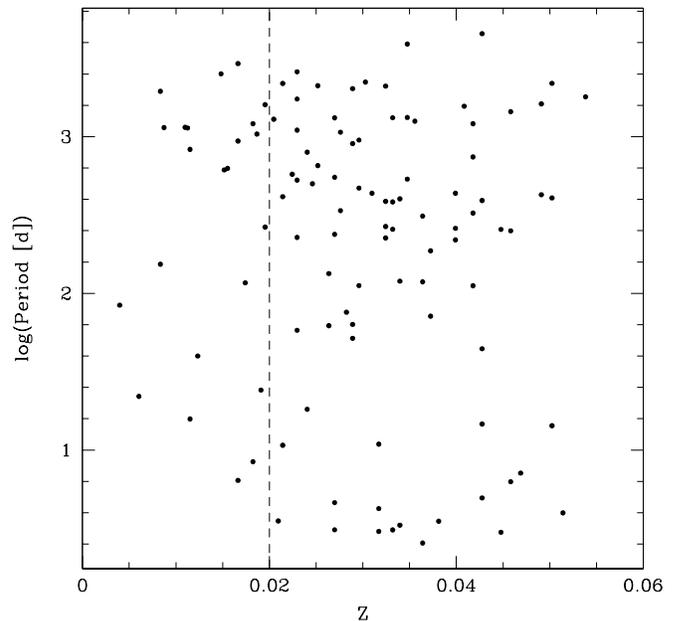}}
\caption{\label{fig:period}
Same as Fig.~\ref{fig:eccen}, but for period $P$ vs.\ metallicity $Z$.
}\end{figure}

If the two mechanisms generated planets with substantially different
distributions in eccentricity, period, or mass, then these should be
revealed in the observed distributions of these properties as functions
of metallicity.  That is, the planets with $Z<0.02$ should entirely
reflect one distribution, while those with $Z\geq 0.02$ should
predominantly reflect the other.

\begin{figure}[b]
\centerline{\epsfxsize=3.5truein\epsffile{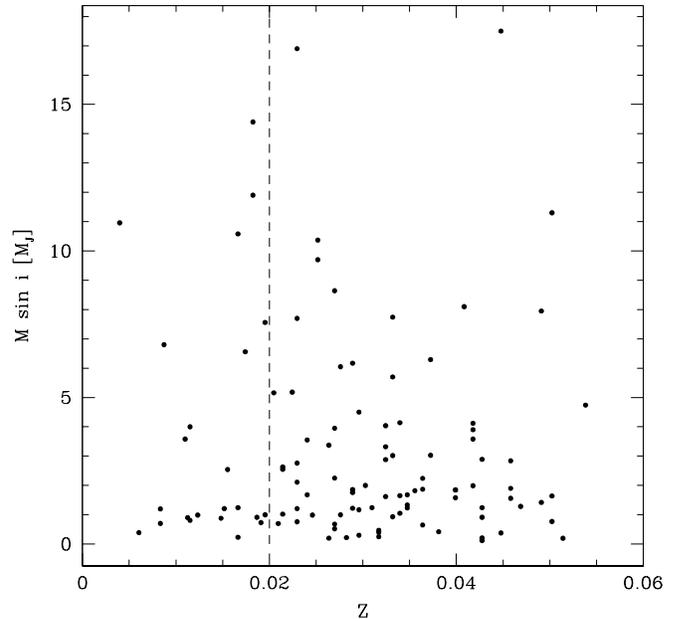}}
\caption{\label{fig:msini}
Same as Fig.~\ref{fig:eccen}, but for mass function $M\sin i$ vs.\
metallicity $Z$.  }\end{figure}

To test this possibility, we make three plots.  In
Figures~\ref{fig:eccen}, \ref{fig:period}, and \ref{fig:msini}, we show
the eccentricities $e$, periods $P$, and mass functions $M\sin i$ of the
109 planets cataloged by \citet{santos} versus the metallicities of
their 97 host stars.  We exclude HD 47536 as its planet's properties are
not well established.  We take planet properties from The Geneva
Extrasolar Planet Search Programmes
website\footnote{http://obswww.unige.ch/$\sim$udry/planet/planet.html}.

No strong pattern emerges from any of these three plots.  The low-$Z$
stars show a modest deficit of short-period planets, but a 
Kolmogorov-Smirnov (KS) test shows a significance of only $p=0.17$.
Similarly, the low-$Z$ stars show a modest deficit of high-mass planets,
but this is even less significant ($p=0.21$).  However,
because only 23 of the 109 planets have metallicities $Z<0.02$, these
plots and tests would only be sensitive to the most glaring differences 
between the two putative populations.

Since there are marginal signals in each of Figures~\ref{fig:period}
and \ref{fig:msini}, we have also plotted (but do not show)
$P$ vs.\ $M\sin i$, with different symbols for the planets of stars
above and below $Z=0.02$.  However, we find no enhanced signal in this
plot, nor indeed in similar plots of $e$ vs.\ $P$ and $e$ vs.\ $M\sin i$.

We can also check whether planetary multiplicity is bimodal in character.
Of the 75 planet-bearing stars with $Z\geq 0.02$, 9 have more than
one planet.  If multiplicity were independent of metallicity, we would
expect 2.64 of the 22 planet-bearing stars with $Z< 0.02$ to have
multiple planets.  In fact, there is only one such metal-poor multiple
system.  Again, however, no strong significance can be attached to this
difference because, based on Poisson statistics, such a shortfall would
be expected 26\% of the time.

Now that a clear break point has been established in planet frequency at
$Z=0.02$, it is critical to establish a larger sample of planets of
sub-solar metallicity stars.  The relatively low frequency of these
planets will make this search difficult, partly because a large number
of such stars must be searched and partly because, as a consequence of
this, these stars will tend to be farther away and so fainter.
Nevertheless, since a comparison of the planet properties of host stars
above and below $Z=0.02$ is the simplest test for bimodal planet
formation, the additional effort required to enhance the low-metallicity
sample would be well justified.

\acknowledgments
This work was supported by grant AST 02-01266 from the NSF.


\end{document}